\documentclass[letterpaper]{article} % DO NOT CHANGE THIS
\usepackage{aaai23}  % DO NOT CHANGE THIS
\usepackage{times}  % DO NOT CHANGE THIS
\usepackage{helvet}  % DO NOT CHANGE THIS
\usepackage{courier}  % DO NOT CHANGE THIS
\usepackage[hyphens]{url}  % DO NOT CHANGE THIS
\usepackage{graphicx} % DO NOT CHANGE THIS
\usepackage{natbib}  % DO NOT CHANGE THIS AND DO NOT ADD ANY OPTIONS TO IT
\usepackage{caption} % DO NOT CHANGE THIS AND DO NOT ADD ANY OPTIONS TO IT
\usepackage{algorithm}
\usepackage{algorithmic}

\usepackage{newfloat}
\usepackage{listings}
\DeclareCaptionStyle{ruled}{labelfont=normalfont,labelsep=colon,strut=off} % DO NOT CHANGE THIS
\floatstyle{ruled}
\newfloat{listing}{tb}{lst}{}
\floatname{listing}{Listing}

\usepackage{times}
\usepackage{helvet}
\usepackage{courier}
\usepackage[normalem]{ulem}
\useunder{\uline}{\ul}{}
\usepackage{multirow}
\usepackage{xcolor}
\usepackage{graphicx}
\usepackage{booktabs}
\usepackage{float}
\usepackage{multicol}
\usepackage{subcaption}
\usepackage{adjustbox}
\usepackage{booktabs}
\usepackage{multirow, makecell}
\usepackage{rotating}
\usepackage{footnote}
\makesavenoteenv{tabular}
\makesavenoteenv{table}
\newcommand{\NofTweet}{20,593,823}
\newcommand{\NofUsers}{7,241}
\definecolor{hotpink}{RGB}{255, 83, 115}
\definecolor{teal}{RGB}{90, 200, 250}
\definecolor{lightgreen}{RGB}{33, 222, 128}
\definecolor{lightblue}{RGB}{72, 123, 232}

\title{Minority Stress Experienced by LGBTQ Online Communities during the COVID-19 Pandemic}
\author {
    Yunhao Yuan\textsuperscript{\rm1}, 
    Gaurav Verma\textsuperscript{\rm2}, 
    Barbara Keller\textsuperscript{\rm1}, 
    Talayeh Aledavood\textsuperscript{\rm1}
}
\affiliations {
    \textsuperscript{\rm1} Aalto University\\
    \textsuperscript{\rm2} Georgia Institute of Technology\\
    yunhao.yuan@aalto.fi, gverma@gatech.edu, barbara.keller@aalto.fi, talayeh.aledavood@aalto.fi
}
\begin{document}
\maketitle
\begin{abstract}
The COVID-19 pandemic has disproportionately impacted the lives of minorities, such as members of the LGBTQ community (lesbian, gay, bisexual, transgender, and queer) due to pre-existing social disadvantages and health disparities. Although extensive research has been carried out on the impact of the COVID-19 pandemic on different aspects of the general population's lives, few studies are focused on the LGBTQ population. In this paper, we develop and evaluate two sets of machine learning classifiers using a pre-pandemic and a during-pandemic dataset to identify Twitter posts exhibiting minority stress, which is a unique pressure faced by the members of the LGBTQ population due to their sexual and gender identities. We demonstrate that our best pre- and during-pandemic models show strong and stable performance for detecting posts that contain minority stress. We investigate the linguistic differences in minority stress posts across pre- and during-pandemic periods. We find that anger words are strongly associated with minority stress during the COVID-19 pandemic. We explore the impact of the pandemic on the emotional states of the LGBTQ population by adopting propensity score-based matching to perform a causal analysis. The results show that the LGBTQ population have a greater increase in the usage of cognitive words and worsened observable attribute in the usage of positive emotion words than the group of the general population with similar pre-pandemic behavioral attributes. Our findings have implications for the public health domain and policy-makers to provide adequate support, especially with respect to mental health, to the LGBTQ population during future crises.
\end{abstract}

\section{Introduction}
The onset of the COVID-19 pandemic has profoundly impacted health, economy, and societies as a whole. It leads to increased anxiety, depression, and stress as people cope with fear of contagion, and uncertainty about the future. Across the world, a wide range of substantive strategies, such as self-isolation, social distancing, and school closures, have been implemented to mitigate the spread of the coronavirus. These policies, which effectively slow the spread of the virus, have exacerbated mental health issues for individuals and communities. 

Isolation from social relationships and disengagement from social activities can lead to adverse mental health outcomes, such as depression, anxiety, and loneliness~\cite{pietrabissa2020psychological}. Recent studies on the effects of the COVID-19 pandemic show a considerable increase in psychological expressions, including expressions related to mental health issues and support, among the general population~\cite{saha2020psychosocial,gallagher2020impact}. The authors highlight major concerns about personal challenges, health issues, and uncertainties surrounding the pandemic. The adverse effects of these policy interventions are shown to have a disproportional impact on different groups and communities within society. For example, recent work by~\citeauthor{pierce2020mental}~(\citeyear{pierce2020mental}) assesses the level of mental distress in the UK population before and during the lockdowns. Their research indicates deteriorated mental health during the pandemic, especially among women, youth, and children.

Due to pre-existing social and structural inequalities, the COVID-19 pandemic is expected to have a greater impact on minority groups, including lesbian, gay, bisexual, transgender, and queer (LGBTQ) individuals. Even in the best of times, LGBTQ community members have experienced worse mental health outcomes resulting from persistent discrimination and stigmatization~\cite{wang2020health}. This community struggles with health disparities, such as higher occurrences of long-term chronic illnesses and unique healthcare needs, compared to the straight and cisgender population~\cite{zeeman2019review}. During the COVID-19 pandemic,~\citeauthor{kneale2020mental}~(\citeyear{kneale2020mental}) report an increased level of stress and depressive symptoms among the LGBTQ community, particularly among transgender people. It is suggested that the closing of educational institutions and mandatory isolation during the pandemic may expose young members of the LGBTQ community to verbal abuse and domestic violence~\cite{salerno2020lgbtq}. However, research to date mainly relies on diagnostic reports and psychological questionnaires to analyze the mental health of the LGBTQ population. Due to the fear of cultural discrimination and stigmatization, a large number of LGBTQ individuals are hard to reach, limiting generalizability. To fill this gap, our study aims to investigate the mental health of the LGBTQ population during the pandemic by leveraging a large sample of social media data.

% social media and it's  psychological impact
%In the wake of the COVID-19 pandemic and the mental health impacts on the LGBTQ individuals that follows, it is crucial to build a deeper understanding of this topic. 
Social media are known to provide spaces where individuals and communities with stigmatized identities can express themselves, talk about their experiences, and seek support, with less fear of offline harm~\cite{andalibi2016understanding}. Data collected from different social media platforms have already been used to investigate stigmatized experiences in various domains such as drug misuse~\cite{garg2021detecting}, sexual abuse~\cite{andalibi2016understanding}, and mental health~\cite{coppersmith2018natural}. Similarly, social media offers an easily accessible and semi-anonymous avenue for LGBTQ individuals to connect with others, seek support and share their concerns~\cite{cannon2017transition}. Consequently, social media data provide an unparalleled opportunity to study the concerns of the LGBTQ community and their evolution over time. Harnessing this opportunity is especially important in the context of the COVID-19 pandemic, when mental health issues are known to have grown significantly among different populations.

% what we do
In this work, we investigate the impact of COVID-19 on the LGBTQ community in the United States via Twitter data. We study \textit{minority stress}, which is defined as the chronic stress derived from stigmatized social status in the context of a heterosexist society faced by the LGBTQ community members due to their sexual and gender identities~\cite{meyer1995minority}.  We address the following three research questions to better understand shifts in the minority stress patterns before and after the breakout of the COVID-19 pandemic and its disproportionate impacts on the LGBTQ population as compared to the general population:

\vspace{0.05in} \noindent\textbf{RQ1}: \textit{How can we measure minority stress on social media using computational methods?}

\vspace{0.05in} \noindent\textbf{RQ2}: \textit{Are there linguistic differences in Twitter posts (tweets) containing minority stress in the time prior to the pandemic compared to minority stress posts during the pandemic?}

\vspace{0.05in} \noindent\textbf{RQ3}: \textit{Are there linguistic differences in tweets between the LGBTQ and the general population during the pandemic?}\vspace{0.05in}

%\gaurav{GV to add a small para here explaining our approach to the above RQs and leave pointers for adding results}

To answer these research questions, we collect \NofTweet~tweets from \NofUsers~US-based LGBTQ Twitter users who disclose their identifies on Twitter from June 2018 to June 2021. We manually annotate 2,800 tweets and divide them into two datasets. One set contains tweets posted before the COVID-19 pandemic, and the other set includes tweets posted during the pandemic. We train multiple machine learning classifiers separately on pre- and during-pandemic datasets, and evaluate the best-performing classifiers on different feature sets. Our best pre-pandemic and during-pandemic models show stable and robust performance with an area under the receiver operating characteristic (AUC) curve of $0.838$ and $0.845$, respectively. We investigate the top statistically significant features of the pre-pandemic and during-pandemic models to compare the emotions behind the minority stress posts on Twitter. We find that anger has become strongly associated with minority stress during the COVID-19 pandemic. Lastly, we apply a stratified propensity score analysis to compare the COVID-19 pandemic effects on the LGBTQ population and the general population on social media. We observe that the COVID-19 pandemic has resulted in a higher usage of cognitive words and a reduced usage of positive emotion words for the LGBTQ population. 

%\gaurav{tense is inconsistent here: remove and paraphrased. We can have -ed throughout the paper consistently, except when discussion the findings, in which case we could use ``we find.''}

\textbf{Privacy and Ethics}
Despite the public accessibility of Twitter data, we adopt several approaches to protect the privacy of the data owners. We remove any information related to personal identity and paraphrase all quotations in this paper to avoid traceability. One of the authors is a member of the LGBTQ community. This helps us better understand the privacy needs and concerns of this community. 

\section{Background and Related Work}
\subsection{Social Media Use During the COVID-19 Pandemic}
Due to the need for social distancing during the COVID-19 pandemic, people increasingly use social media not only to consume and share information, but also to engage in seeking help, advice, and providing supportive information~\cite{saud2020usage}. Recent work by~\citeauthor{tsao2021social}~(\citeyear{tsao2021social}) highlights various facets of public health research on social media during the pandemic, such as psychological reactions, government responses, and public attitudes. 

~\citeauthor{xiang2021modern}~(\citeyear{xiang2021modern}) collect data from Twitter and Decahose to analyze the public discourse and sentiments relative to older people and the extent of ageism among social media users. ~\citeauthor{zhou2021detecting}~(\citeyear{zhou2021detecting}) build a depression classifier by extracting multi-modal features from Twitter data, such as emotion, topic, and domain-specific features, and analyze the local dynamics during the COVID-19 in Australia.~\citeauthor{Verma2022}~(\citeyear{Verma2022}) investigate the 80 million Twitter posts and demonstrate that users sharing COVID-19 related misinformation will experience exacerbated anxiety.
%investigate the multi-linguistic misinformation on various social media and find that politics covers most of the misinformation in the their dataset.

\subsection{LGBTQ Community on Social Media}
% LGBTQ+ individuals' life on different platform
Social media plays a prominent role in the exchange of information. It also provides opportunities and platforms for minorities and communities, such as LGBTQ individuals, to engage in online activities, such as seeking out social support, exchanging information, and maintaining social relationships~\cite{steinke2017meeting}. Prior research suggests that LGBTQ youths use social media as a primary method of socializing and spend more time online than their counterpart cisgender individuals~\cite{steinke2017meeting}. 

% LGBTQ+ researches on social media
Given the significant growth in social media usage, data from LGBTQ online communities has been used to analyze social media's effects on the LGBTQ population. One study~\cite{craig2014you} illustrates that LGBTQ youth use social media platforms as a tool for actively coming out as they are perceived as safe places. In another study, researchers demonstrate the feasibility of evaluating equality of hospital care for LGBT individuals through analyzing online discussions related to LGBT hospital care on Twitter~\cite{hswen2018investigating}.~\citeauthor{haimson2019mapping}~(\citeyear{haimson2019mapping}) show that the positive sentiment increases over time on average after disclosing their identities to people with close relationships by analyzing posts from Tumblr. Overall, existing research indicates a need to understand better the relationship between the LGBTQ community and their use of social media.

\subsection{Minority Stress Theory}
The minority stress theory~\cite{meyer1995minority} provides a full conceptual framework for understanding how minority stressors impact the mental health of LGBTQ individuals. It defines minority stressors as chronic stress derived from stigmatized social status in the context of a heterosexist society.~\citeauthor{meyer1995minority}~(\citeyear{meyer1995minority}) suggests three types of minority stressors: internalized homophobia, perceived stigma, and prejudice events. Stressors specific to LGBTQ individuals typically contain social stigma, prejudice, and discrimination, such as verbal abuse, harassment, and physical violence.

The minority stress theory has been utilized as a theoretical framework for numerous studies to examine the mental health of gender and sexual minorities. Previous studies have found widespread mental health disparities between the LGBTQ and the broader population in generalized anxiety symptoms~\cite{borgogna2019anxiety}, depression~\cite{kulick2017heterosexism}, and substance abuse~\cite{goldbach2017sexual}.~\citeauthor{lehavot2011impact}~(\citeyear{lehavot2011impact}) analyze the links of minority stress with mental health and substance use among women belonging to a sexual minority.~\citeauthor{frost2015minority}~(\citeyear{frost2015minority}) find that minority stress impacts the physical health of the LGBTQ population. 

\begin{table*}
\centering
\def\arraystretch{1.5}
\begin{tabular}{p{0.15\linewidth}  p{0.8\linewidth} } 
\toprule
\textbf{Category} & \textbf{Tweet Example} \\ 
\midrule
\textbf{Prejudice Events}         
& \textit{I was feeling really good until I was called, and then the verbal threats started once they knew I was trans. I'm at a loss for words.} \\
\textbf{Perceived Stigma}         
& \textit{I believe that my desire to be visibly gay in games is interfering with my enjoyment of the ones I have. I'm sick of modifications stitching things together and half-measures. I just want to play games and not wear a mask of myself.} \\
\textbf{Internalized LGBTQphobia}
& \textit{Being transgender is something I despise. Why couldn't I just be a cisgirl from the start? Every ray of optimism that you have ends up destroying you. Either you let it kill you for not being yourself, or you let the world kill you for being yourself.} \\
\bottomrule
\end{tabular}
\caption{Slightly paraphrased sample tweets that belong to different categories of minority stress. \label{tab:codebook}}
\end{table*}

\section{Data}
We leverage Twitter timeline data of users who disclose their identities on Twitter as LGBTQ members. The data collection includes four steps: 1) curating a list of LGBTQ related hashtags; 2) identifying LGBTQ users; 3) collecting the timeline data of these self-disclosing LGBTQ users; 4) building a control dataset from non-LGBTQ users. We apply a codebook from~\cite{saha2019language} to annotate tweets in the LGBTQ dataset. 
%The following subsections will provide details of these steps.

%We collect data spanning the time between June 1st, 2018 to June 1st, 2021 from Twitter users in the US who identify themselves as being part of the LGBTQ community. The data collection includes three steps: 1) creating a list of LGBTQ related hashtags and collecting tweets containing these hashtags; 2) collecting and filtering the biographies of the authors who posted the LGBTQ relevant tweets; 3) collecting the timeline data of these users who identify themselves as being part of the LGBTQ community, and collecting another set of users who do not self-reportedly belong to the LGBTQ community as a control group. The following subsections will provide details of these steps. Next, we apply~\citeauthor{saha2019language}~(\citeyear{saha2019language}) codebook of the minority stress into our dataset and use the codebook for annotation.

\subsubsection{LGBTQ Related Hashtag List}
The first step in this process is seeding our sample with two popular LGBTQ relevant hashtags, \#gay and \#lgbt. We use the Twitter Application Programming Interface (API) to collect the latest 10,000 tweets posted in the US for each seeding hashtag and rank the top 100 co-occurring hashtags by their frequencies. By manually examining the hashtags and removing those with wide-ranging or ambiguous meanings, we find 60 LGBTQ relevant hashtags (Full list in Supplemental Material). After collecting 10,000 tweets containing at least one of the LGBTQ relevant hashtags, we get a corpus of 600,000 LGBTQ relevant tweets.

%For our data collection, we use the Twitter Academic API to obtain tweets posted from June 1st, 2018 to June 1st, 2021, and the biographies \gaurav{biographies in place of biographies; multiple instances} of their authors. We look for signals that can tell whether the users are LGBTQ individuals or cisgender individuals from both the tweets and the user bio description. The first step in this process is seeding our sample with the two popular LGBTQ relevant hashtags, \#gay and \#lgbt. We collected 10,000 tweets posted in the US for each of the hashtags and ranked the top 100 co-occurring hashtags by their frequencies. By manually examining the hashtags and removing hashtags with wide-ranging or ambiguous meanings, we found 60 LGBTQ relevant hashtags. This process resulted in a corpus of 600,000 LGBTQ relevant tweets.

\subsubsection{Twitter Data of Members of the LGBTQ Community}
We fetch the bios of the accounts which posted these 600,000 LGBTQ relevant tweets. Bios are public descriptions of Twitter accounts with a maximum length of 160 characters, where users typically disclose information about who they are. We search these bios using regular expression matching (e.g., gay, lesbian, and bisexual) and pride-related emojis, such as the rainbow flag, pride flag, and transgender symbol. We retrieve~\NofUsers~distinct accounts belonging to self-disclosing LGBTQ Twitter users. We randomly select 40 users from this set and manually check if their latest 50 tweets explicitly express their identities. Based on manual verification done by the authors, we confirm 32 users to be LGBTQ individuals. The bios of the eight remaining users contain terms with LGBTQ keywords or emojis, which are either supportive expressions (e.g., \textit{LGBTQ life matters}) or declare their relationship with the LGBTQ community (e.g., \textit{LGBTQ* Advisory Committee in College}).

%\barbara{how so? and what is different to the two other users pointed out above?}
\subsubsection{Compiling the LGBTQ Dataset} 
From the \NofUsers~ unique LGBTQ users, we collect Twitter metadata, including the number of tweets, likes, followers, followees, and the account creation time. Inspired by~\citeauthor{pavalanathan2016more}~(\citeyear{pavalanathan2016more}), we filter out untypical users, such as inactive users and advertising accounts by removing users who have more than 5000 followers/followees or who posted fewer than 200 or more than 30,000 tweets. 
For the remaining 5,708 users, we collect their timeline data from June 1st, 2018 to June 1st, 2021. In the end, the \textit{LGBTQ} dataset contains 5,708 users. Among these users, 972 users identify as gay, 173 users as lesbian, 1,424 users as transgender, and 95 users as bisexual. The remaining 3,044 users, do identify as part of the LGBTQ community but are not categorized into one of the former categories.

\subsubsection{Compiling the Control Dataset} 
To compare the linguistic differences between the LGBTQ and the general population during the pandemic, we build a \textit{Control} dataset of users who do not use LGBTQ keywords in their bios. Recall that our RQ3 is to compare the linguistic difference between the general and the LGBTQ population. We seek to isolate the effect of the pandemic on two groups of users with similar pre-pandemic attributes. To find non-LGBTQ users who have similar linguistic patterns to the LGBTQ ones, we rank the top 100 hashtags from all tweets in the LGBTQ dataset and manually remove hashtags that occur in our LGBTQ related hashtag list, which results in 79 hashtags  (Full list in Supplemental Material). For each of these 79 hashtags, we collect 10,000 tweets and bios of their authors. After removing users whose bios contain LGBTQ relevant keywords and/or LGBTQ emojis, we collect the Twitter metadata of the users and limit the dataset using the same methods as for the LGBTQ dataset. We collect the timeline data of the remaining 36,030 users to build the \textit{control} dataset.

\subsection{Annotating Minority Stress on Social Media}
We use a codebook from previous research~\cite{saha2019language} to annotate minority stress on our LGBTQ dataset. The codebook is developed from Meyer's minority stress framework~\cite{meyer1995minority} to identify minority stress in social media. It conceptualizes minority stress as three categories: prejudice events, containing discrimination and violent experiences; stigma, involving expectations of being rejected because of one's own minority identity; and internalized homophobia, related to the negative attitudes toward oneself. Table~\ref{tab:codebook} presents some paraphrased tweet examples classified into these categories of minority stress.

To analyze the linguistic differences in minority stress tweets between the time span before and during the COVID-19 pandemic, it is essential to keep the number of tweets in the training dataset balanced between the two time periods. We randomly selected 2800 tweets for annotation, with 1400 tweets created between June 1st, 2018 and November 30th, 2019, referred to as the \textbf{pre-pandemic dataset}, and 1400 tweets created between 1st December, 2019 and June 1st, 2021, referred to as the \textbf{during-pandemic dataset}.

\subsubsection{Background of the Annotators} 
Four authors of this paper and two recruited annotators participate in the annotation process. The group consists of four females and two males, each with diverse cultural background. Besides, one of the annotators belongs to the sexual minority community, which might help in explaining the highly subjective, intrinsic, and complicated context of the tweets to be annotated and the final goals of the annotation task.

\subsubsection{Annotation Task} 
Four authors independently annotate 200 randomly-sampled tweets and revise the codebook by discussing the discrepancies. Together, the annotators decide only to annotate tweets discussing recent and present experiences pertaining to minority stress. Lacking the context of URLs and the complexity of checking the contents in URLs, the URLs in posts are removed, but the remaining content of the tweet is included in the dataset. %In cases where we have different interpretations for the tweets describing minority stress experienced by others, after a discussion, we agree that this kind of tweet is annotated as minority stress by considering the tweet owner will also feel attacked, discriminated against, or stigmatized from experience by another person as they share the minority identity. 

After establishing the annotation rules, two university students are recruited to annotate the remaining 2,600 tweets. They are asked to learn the codebook with the code explanations, discrepancies, and tweet examples. The two annotators annotate the same sample sets to avoid bias and increase accuracy. In case annotators have any uncertainties about a tweet, they independently discuss it with the authors. The discussion leads to substantial inter-rater reliability, with an overall Cohen’s kappa ($\kappa$) coefficient~\cite{landis1977measurement} of 0.727, which is an improvement over the pre-discussion ($\kappa$) of 0.542.  If there are any remaining tweets with no agreement after discussion, the annotators and authors examine them together, and the final labels are decided by majority voting. Out of the 2,800 annotated tweets, 684 are categorized as prejudice events, 306 as perceived stigma, and 36 tweets as internalized LGBTQ phobia. Minority stress subcategories are not mutually exclusive and tweets do not provide the full information on the context. So, we subsequently focus on the binary categorization of ``minority stress''  and ``non-minority stress''.

\begin{figure*}
	\centering\includegraphics[width=5.5in]{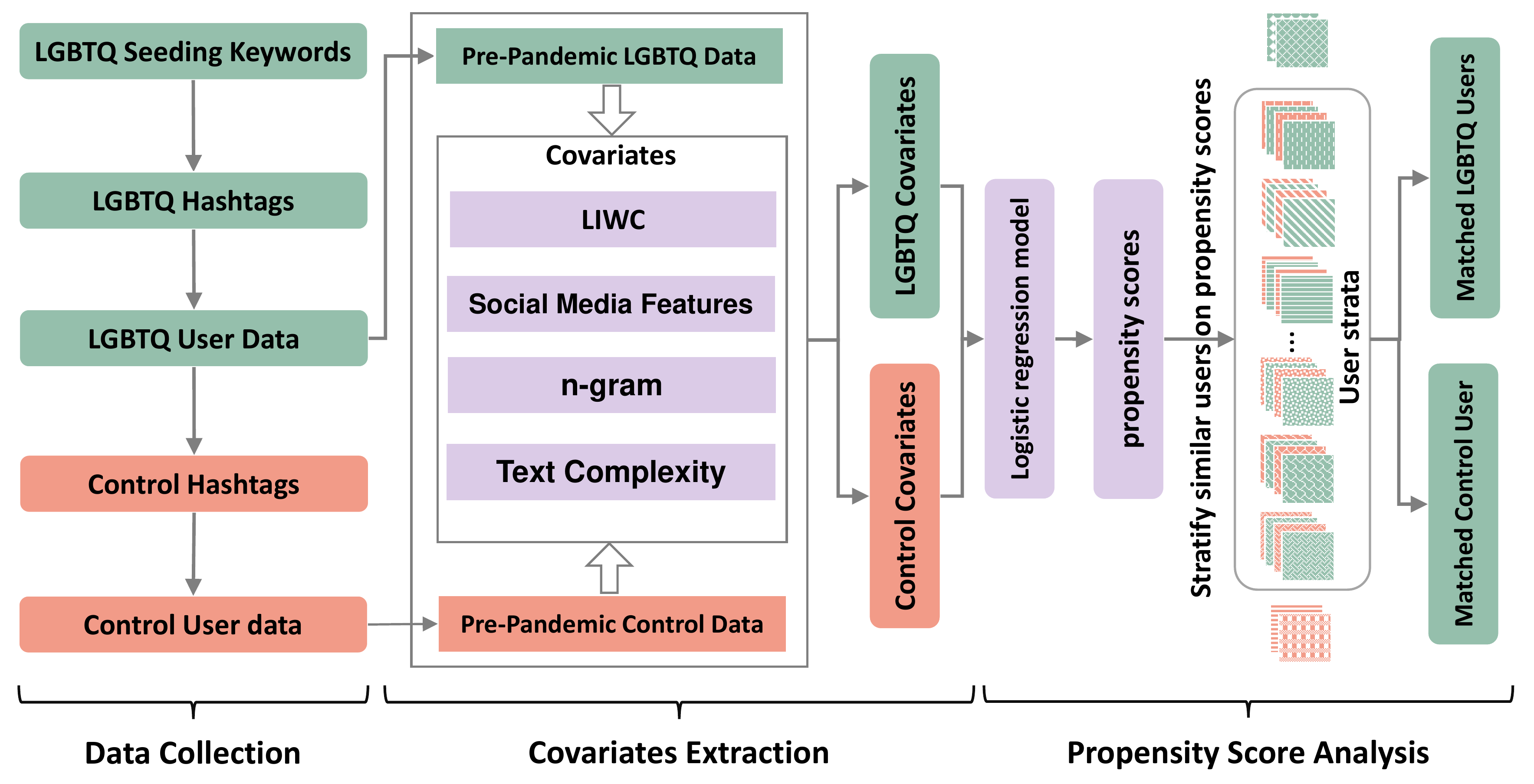}
	\centering
	\caption{Schematic diagram of propensity score matching between LGBTQ users and general users.}
	\label{fig:diagram}
\end{figure*}

\section{Methods}

\subsection{Study Design and Rationale}
Our study aims to answer three research questions --- measuring minority stress using social media posts (RQ1), capturing the linguistic differences in minority stress posts by LGBTQ members before and during the COVID-19 pandemic (RQ2), and contrasting the linguistic difference of LGBTQ members with the general population to explain their emotional reactions to the pandemic (RQ3). For RQ1, we conduct annotations based on the existing taxonomy of minority stress and develop machine learning classifiers using the annotated data. For RQ2, we calculate the coefficient scores for every input feature and analyze the most discriminative features for the best-performing models. RQ3, however, involves additional considerations --- given that the linguistic difference of LGBTQ members can precipitate due to a variety of reasons beyond their sexual and gender identities, it is essential to contrast these differences with carefully chosen members of the general population. By closely matching a series of pre-pandemic attributes between members of the general population and members of the LGBTQ population, we try to isolate the effect of belonging to a minority group on the respective emotional response to the pandemic. To this end, we adopt a causal inference framework~\cite{imbens2015causal} to address RQ3. The schematic diagram of our approach is shown in Figure~\ref{fig:diagram}. Our approach first matches members of the general population (i.e., the \textit{control} group) with members of the LGBTQ group (i.e., the \textit{treatment group\footnote{LGBTQ group: Conventionally, the causal inference terminology uses ``treatment'' to refer to intervention or exposure of interest that is being studied, but we do not insinuate that someone of this group needs treatment or has been treated. In this paper, we use ``LGBTQ group'' to replace the term ``treatment group''.}}) based on pre-pandemic behavioral attributes. For this, we train a machine learning classifier to estimate the likelihood of an individual being assigned to either the LGBTQ or control group (i.e.,\textit{ propensity}) based on covariates and perform matching across groups using estimated propensity scores. Within matched groups of control and LGBTQ groups, we analyze the linguistic difference during the pandemic to compare the behavior of similar individuals who differ in terms of their sexual and/or gender identities. In sum, our approach ensures that members of the LGBTQ population and the general population who are being compared have similar behaviors prior to the pandemic. This gives us the means to analyze the linguistic differences between LGBTQ members, compared to the members of the general population.

\subsection{Machine Learning Classifier for Evaluating Minority Stress}
Our methodological approach describes the way that we conceptualize the intrinsic language features within tweets and apply them to a series of machine learning models. 
\subsubsection{Feature Engineering}
Each tweet in the dataset is described as a vector of the following feature sets:
\begin{itemize}

    \item \textbf{Latent Semantics (Word Embeddings)}: To capture language semantics beyond raw keywords, we use word embeddings, which are essentially vector representations of words in latent semantic dimensions. In particular, we use pre-trained word embeddings (GloVe from StanfordNLP~\cite{pennington2014glove}) in 50 dimensions that are trained on word-word co-occurrences in a Wikipedia corpus of 6B tokens.
    
    \item \textbf{Psycholinguistic Attributes (LIWC):} The use of Linguistic Inquiry and Word Count lexicon~\cite{tausczik2010psychological} is a well-established approach in extracting psycholinguistic features of texts in multiple dimensions, consisting of affection, cognition and perception, interpersonal focus, temporal references, lexical density and awareness, biological concerns, and social and personal concerns~\cite{de2014mental}.
    
    \item \textbf{Sentiment:} Stanford CoreNLP’s deep learning based sentiment analysis tool~\cite{manning2014stanford} is used to measure the sentiment of the tweets. We use this tool to calculate the positive, negative, and neutral scores for each tweet.
    
    \item \textbf{Open Vocabulary (n-grams):} Open Vocabulary has been used in a previous study~\cite{kim2021you} to examine the psychological features of individuals. Similar to their approach, we apply top 500 uni- and bi-grams as open vocabulary features in our dataset.

\end{itemize}

\subsubsection{Machine Learning Models}
Using the pre-pandemic and during-pandemic datasets, we build two sets of binary classifiers for detecting tweets with minority stress. We refer to the models trained on the pre-pandemic dataset as \textbf{pre-pandemic models} and the ones trained on the during-pandemic dataset as \textbf{during-pandemic models}.

Similar to \cite{saha2019language}, we implement and evaluate six models for each training dataset, including a dummy classifier (baseline), na\"ive Bayes, Logistic Regression, Decision Tree, Support Vector Machine (SVM), and Multilayer Perceptron (MLP) algorithm. For hyperparameter tuning, all classifiers go through $k$-fold cross-validation ($k = 10$).

\subsubsection{Feature Importance}
To understand which features contribute the most to classification, we fit the best-performing model to the two training datasets (pre-pandemic and during-pandemic) separately. It turns out that logistic regression models show the best performance. We, therefore, retrieve coefficient scores for each input variable and compare their importance. In the Results section, we focus on the interpretable feature sets, namely LIWC, and n-grams. These two feature sets turn out to be the most important feature sets across all feature sets to predict minority stress. We rank all features with their weights. Features with positive weights contribute to identifying the minority stress tweets, while features with negative weights contribute to the opposite decision of ``no minority stress''. 
%As we rank the weights, this results in the features with the highest rank if they have negative weight, contributing to the "No minority stress" decision.
%Features with higher rankings indicate the better performance to identify tweets with the minority stress, while features with lower rankings indicate the stronger ability to classify non-minority stress tweets. 

\subsection{Matching For Causal Inference} 
\subsubsection{Matching Covariates} 
Our goal is to compare the linguistic difference in tweets between members of the LGBTQ population and the general population during the COVID-19 pandemic. To minimize the possible occurrence of selection bias, matching is an efficient strategy in case-control studies to estimate causal effects. In our case, we build several covariates from the LGBTQ and control datasets using data before December 1st, 2019, to control for similar pre-pandemic behavior on social media. The first set of covariates comprises Twitter users' social media features (the number of tweets, likes, followers, followees, account creation time, and posting frequency). The second set contains the distribution of word usage in Twitter timelines. We extract the top 500 unigrams as the second covariates set. The third set includes the psycholinguistic features in the timeline data by measuring the word distribution in the LIWC lexicon~\cite{wei2021linguistic}. As previous research has indicated that mental health is correlated with text readability~\cite{tausczik2010psychological}, we add several metrics of readability into the last set.

\subsubsection{Propensity Score Analysis} 
To ensure the similarities between our LGBTQ dataset and the Control dataset, matching is used to pair LGBTQ users and control users, whose covariates are similar to each other. We implement a logistic regression classifier to predict the likelihood of a user belonging to the LGBTQ group or control group based on their covariates. We refer to the likelihood as the propensity score. 

\begin{figure}[!t]
\subfloat[\label{sfig:a_matching}]{%
  \includegraphics[width=.495\linewidth]{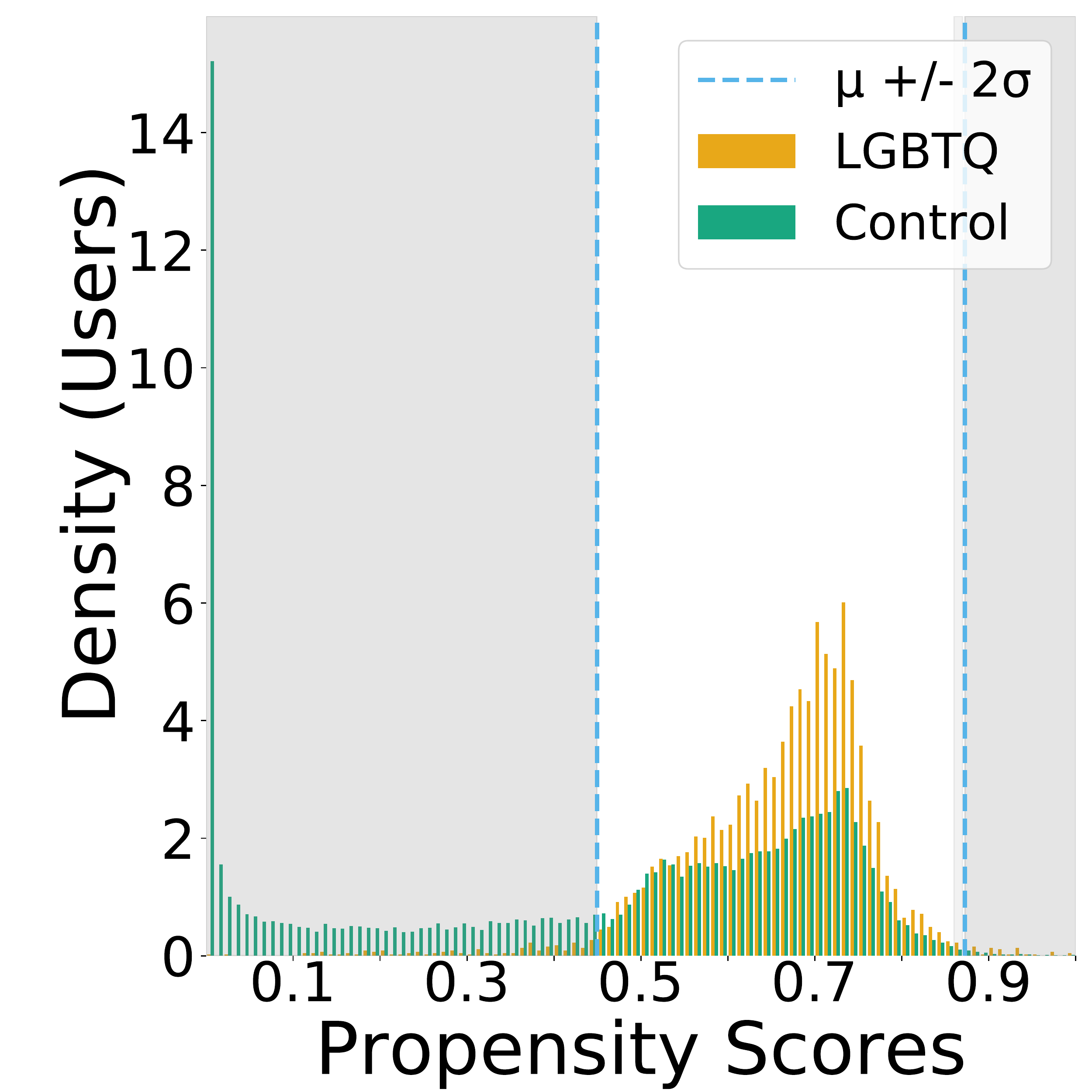}%
}
\subfloat[\label{sfig:b_matching}]{%
  \includegraphics[width=.495\linewidth]{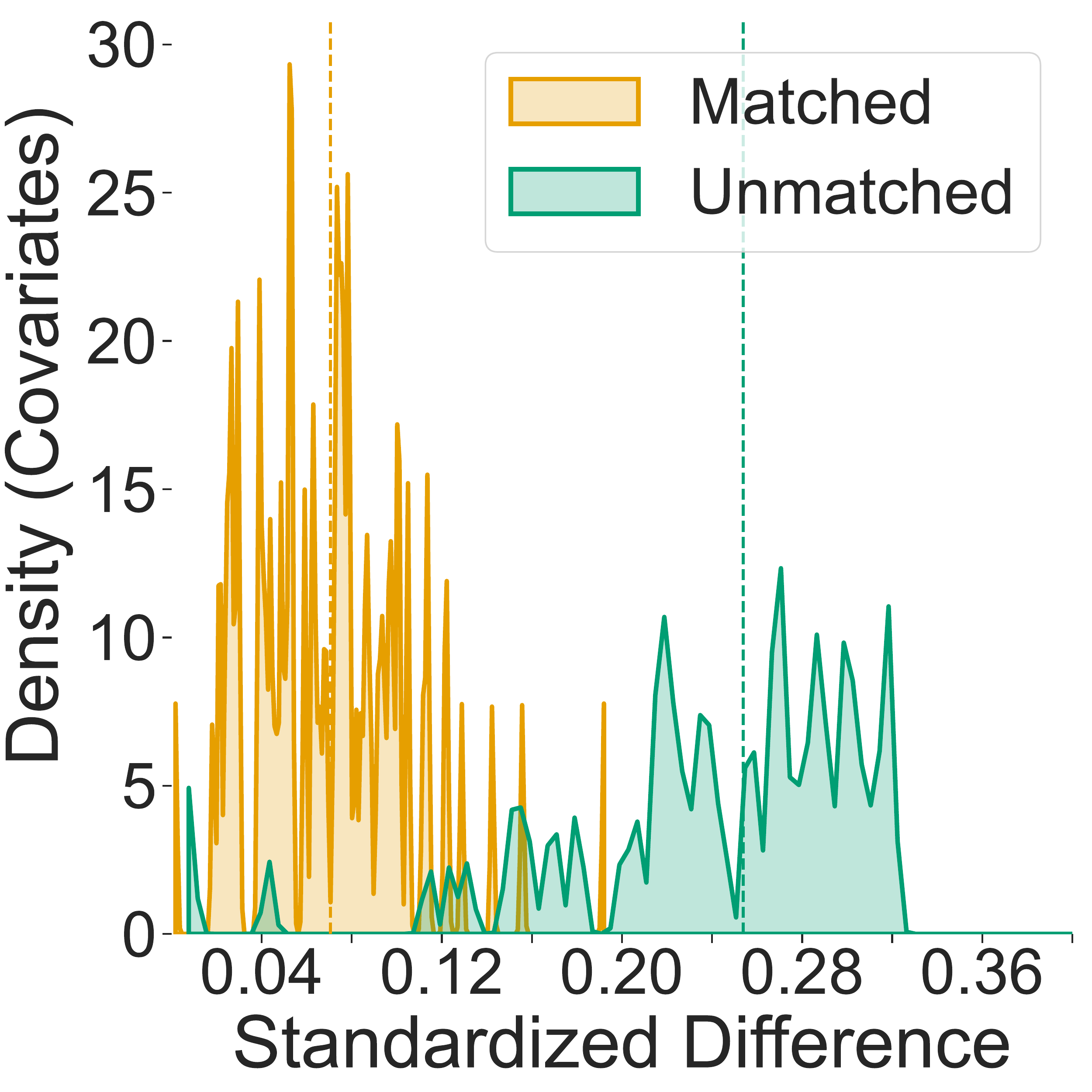}%
}\hfill

\caption{ (a) Propensity score distribution (shaded region are the dropped strata), (b) Quality of matching.}
\label{fig:matching}
\end{figure}

We divide the propensity score distribution into 100 strata with equal width. The users with similar propensity scores are grouped into the same stratum~\cite{kiciman2018using}. This helps us evaluate the effects of the COVID-19 pandemic within each stratum, where the control group users are matched to the LGBTQ users based on the pre-pandemic behavioral traits. We remove the users with propensity scores falling within two standard deviations from the mean. We drop the strata failing to satisfy the minimum sample size within each stratum based on previous causal inference research~\cite{de2016discovering}. By ensuring that there are at least 50 users per group in each stratum, this approach results in 33 strata, containing 4,107 LGBTQ and 7,129 Control users (Figure \ref{sfig:a_matching}).

\subsubsection{Quality Assessment of Covariate Matching}
To assess whether users in the LGBTQ and control groups are statistically comparable, we measure the balance of the covariates. Comparisons between the two groups in each of the 33 valid strata are made by calculating the standardized mean difference (SMD). SMD is the difference in the mean covariate values between the LGBTQ and control groups divided by the pooled standard deviation of the two groups. A previous study~\cite{kiciman2018using} has indicated that two groups are balanced if the absolute SMD of all covariates is less than 0.2. For the unmatched dataset, the maximum absolute SMD is 0.8, and the mean is 0.25, while the maximum absolute SMD is 0.19 and the mean is 0.07 in the matched dataset (Figure \ref{sfig:b_matching}). This satisfies the suggested threshold of less than 0.2.

\subsubsection{Estimating the Average Treatment Effects}
To estimate the effect of the COVID-19 pandemic, we compute the change in word usage in the LIWC variables. The treatment effect\footnote{treatment effect: The phrase ``treatment effect' is used in causal inference to describe the causal impact of a specific intervention or exposure that is being investigated. It's important to note that we do not intend to imply that individuals in the studied group require or have received any form of treatment.}(TE) of an outcome variable $S$ is calculated in stratum $k$ by the following equation:
\begin{equation}
TE_{k} = \frac{\sum_{i}^{N_{k}}(S^{\small{\textrm{during}}}_{L_i} - S^{\small{\textrm{before}}}_{L_i})}{N_{k}} -  \frac{\sum_{j}^{M_{k}}(S^{\small{\textrm{during}}}_{C_j} - S^{\small{\textrm{before}}}_{C_j})}{M_{k}},
\end{equation}
where $N_{k}$ is the number of LGBTQ users in stratum $k$, and $M_{k}$ is the number of matched control users in the same stratum. $S^{\textrm{before}}_{L_i}$ and $S^{\textrm{before}}_{C_j}$ are the word usage in the LIWC variables measured for LGBTQ user $L_i$ and control user $C_j$ before the pandemic. $S^{\textrm{during}}_{L_i}$ and $S^{\textrm{during}}_{C_j}$ are the word usage in the LIWC variables measured for LGBTQ user $L_i$ and control user $C_j$ during the pandemic. After calculating the average TE across all strata, we obtain the mean TE for the pandemic per LIWC variable. The outcome is interpreted as a higher increase  (greater than 0) or higher decrease (less than 0) of observable attributes during the pandemic of LGBTQ users compared to the control group with similar pre-pandemic attributes.
%The outcome is interpreted as an better or worse observable attributes in word usage during the pandemic of LGBTQ users, compared to the control group with similar pre-pandemic attributes. 

\section{Results}
\begin{figure}[!t]
\subfloat[Pre-pandemic model \label{sfig:a_CMatrices_a}]{%
  \includegraphics[width=.48\linewidth]{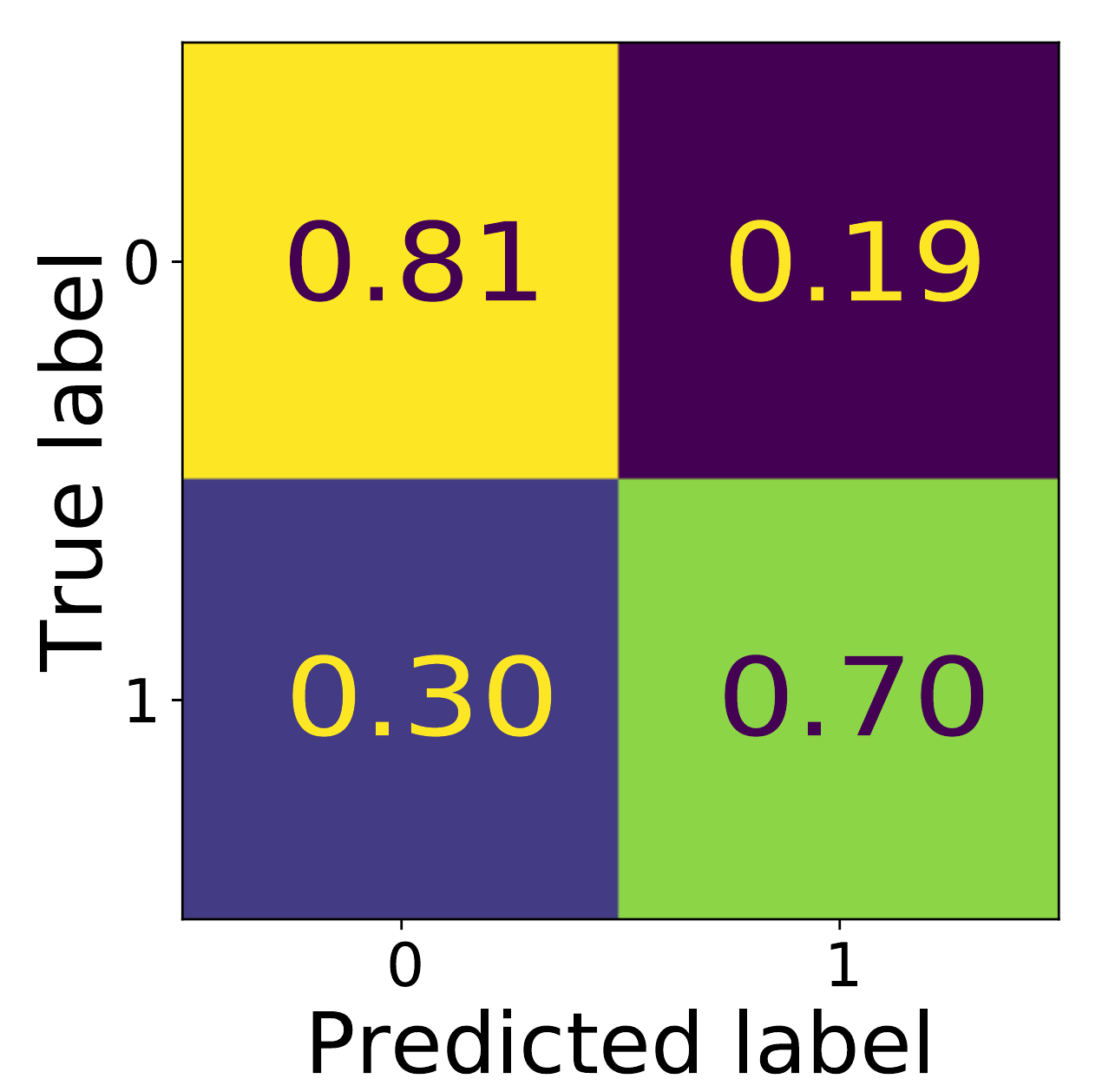}%
}
\subfloat[During-pandemic model \label{sfig:b_CMatrices_b}]{%
  \includegraphics[width=.48\linewidth]{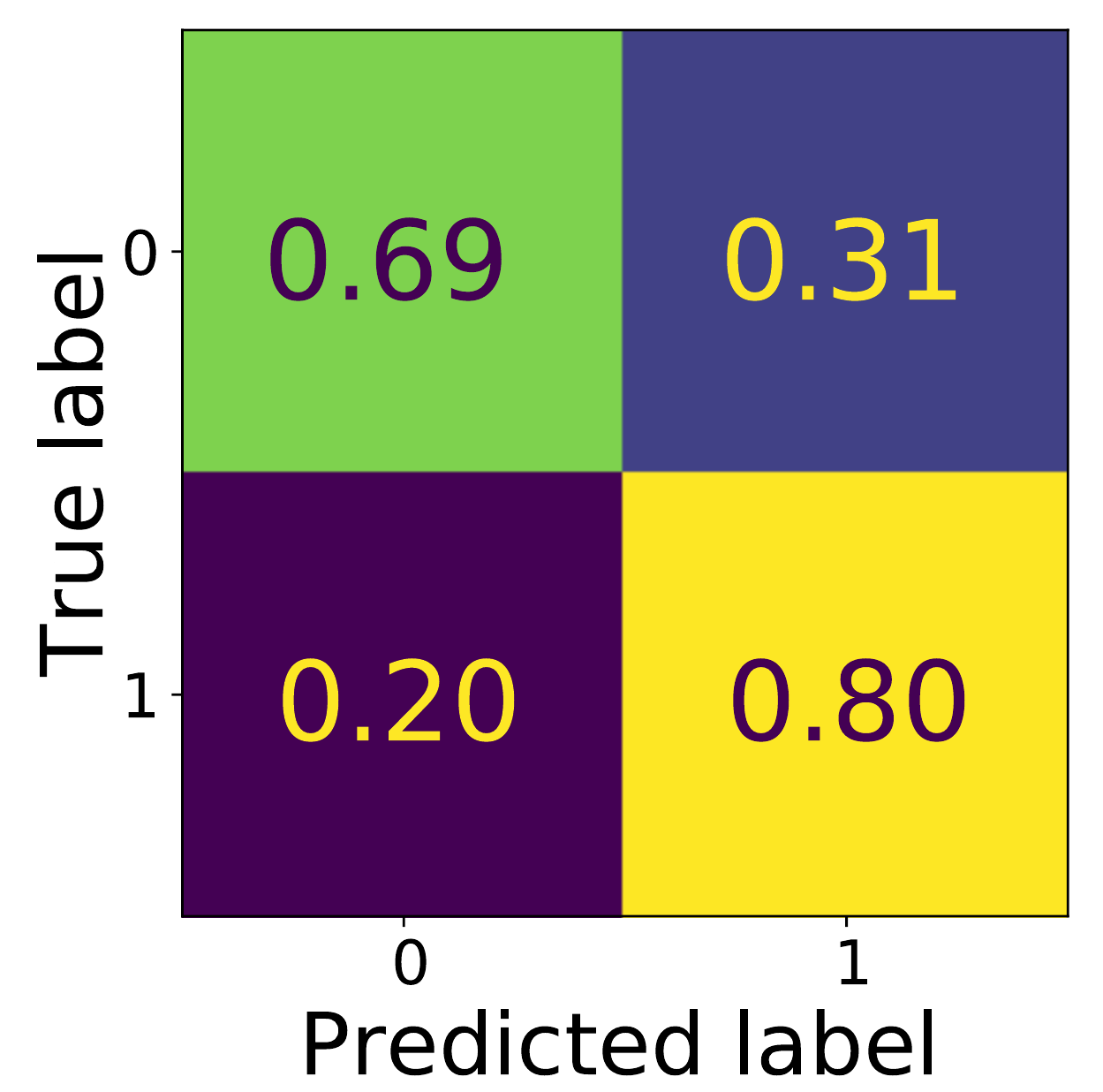}%
}
\caption{ (a) Confusion matrices of the best-performing pre-pandemic model, (b) Confusion matrices of the best-performing during-pandemic model.}
\label{fig:CMatrices}
\end{figure}

\begin{figure}[!t]
\subfloat[Pre-pandemic model\label{sfig:c_ROC}]{%
  \includegraphics[width=.49\linewidth]{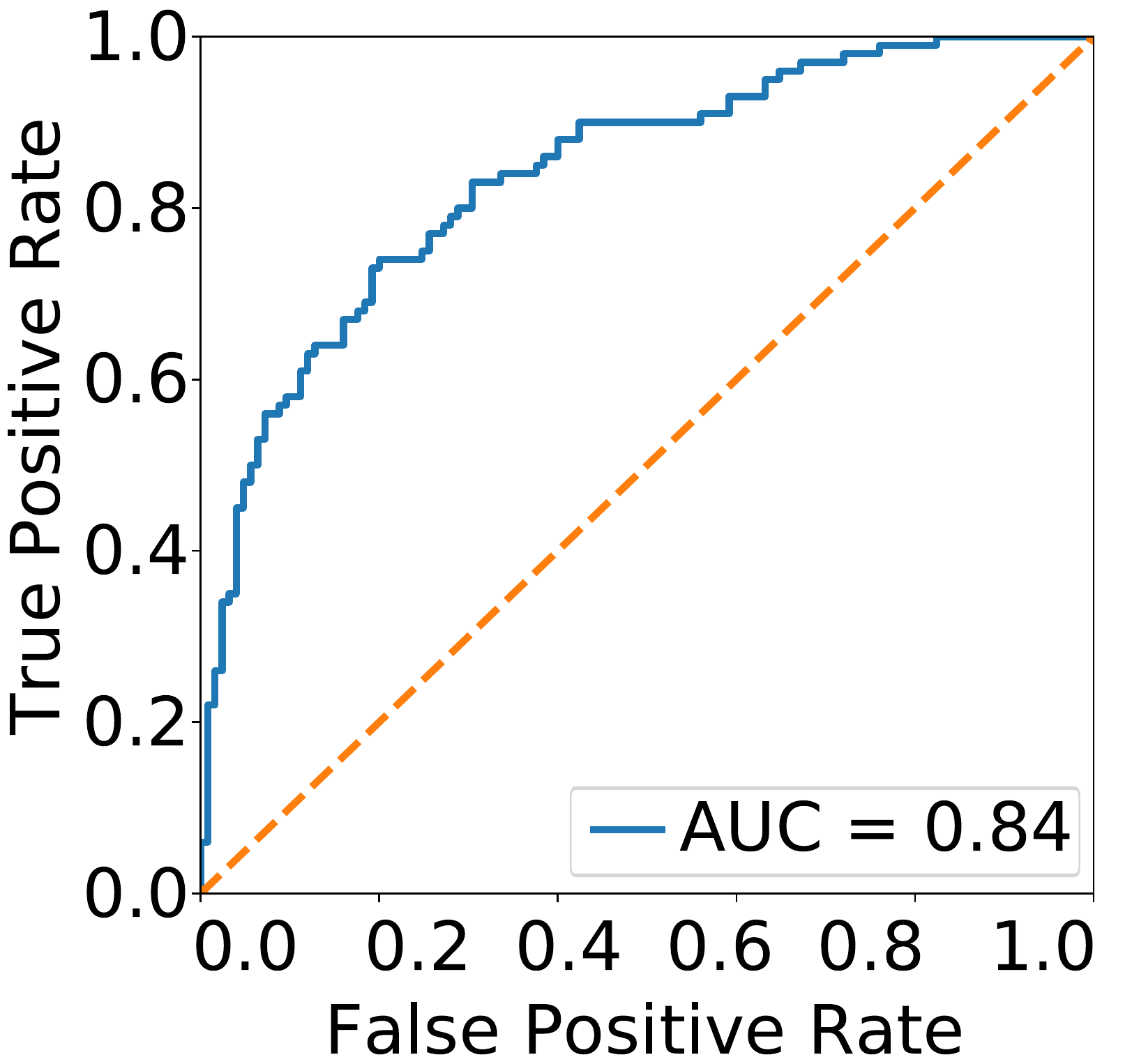}%
}
\subfloat[During-pandemic model\label{sfig:d_ROC}]{%
  \includegraphics[width=.485\linewidth]{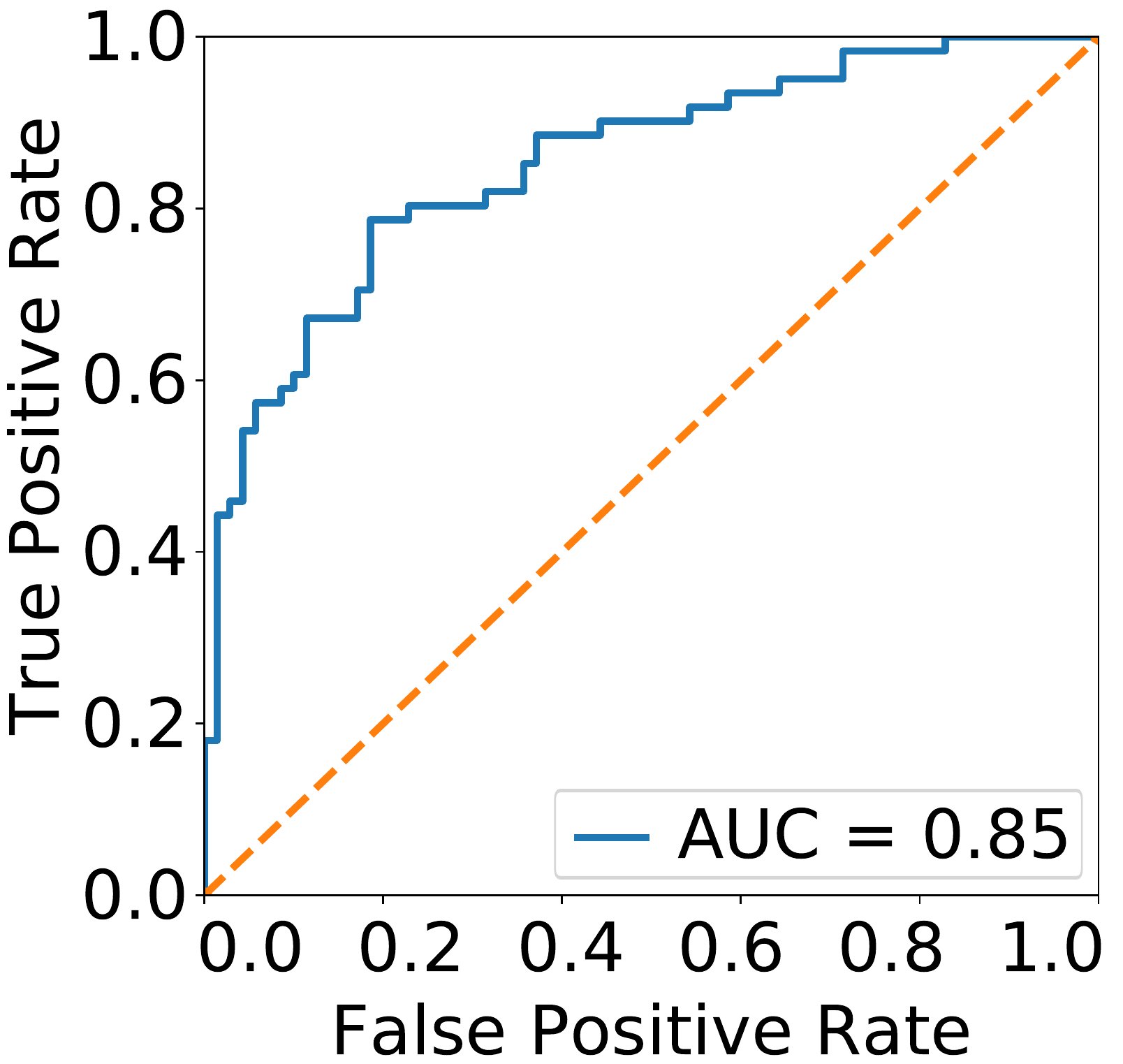}%
}
\caption{(a)  ROC curves of the best-performing pre-pandemic-model (b) ROC curves of the best-performing during-pandemic model.}
\label{fig:ROC}
\end{figure}

In this section, we examine if different feature sets affect the preference of the models (RQ1). We investigate the linguistic difference before and during the pandemic in the context of minority stress (RQ2). We compare the linguistic differences in tweets during the COVID-19 pandemic between the general and the LGBTQ population (RQ3). 

\subsection{RQ1: Machine Learning Classifier Performance}
We implement and evaluate multiple classifiers in the pre-pandemic training dataset and during-pandemic training dataset, including a dummy classifier (baseline), logistic regression, random forest, support vector machine, na\"ive bayes, and multilayer perceptron algorithm. Table \ref{tab:modelperformance1} summarizes the performance of our models and includes the average precision, recall, F1, accuracy, and AUC. We find that logistic regression models show strong and stable performance to indicate minority stress. The best pre-pandemic model shows 75.3\% accuracy with an AUC of 0.838, while the best during-pandemic model shows 76.5\% accuracy with AUC of 0.845, which are used for further analysis. 

We investigate which feature sets in practice contribute the most to the preference of our classifier. We evaluate our model under five different feature sets: a full feature set as well as each of the four sets excluding one of the feature sets. Table \ref{tab:modelperformance2}  summarizes the performance of each feature set in logistic regression models trained on pre-pandemic and during-pandemic datasets. We find that all pre-pandemic and during-pandemic models perform considerably better than the baseline. This reveals that our feature sets do adequately capture the minority stress in tweets, and these different features in the training set do boost the performance of machine learning classification. In addition, we generate the confusion matrix (Figure \ref{fig:CMatrices}) and the receiver operating characteristic (ROC) curves (Figure \ref{fig:ROC}) of the best-performing models. We see that the during-pandemic model shows a higher false positive score than a false negative score, but the pre-pandemic model has the opposite.

%\gaurav{so far we have been saying `we analyze this and the results are there' – however, we should include a line or two on what we infer from the presented results.}

\begin{table}[!t]
\centering
\resizebox{0.45\textwidth}{!}{%
\begin{tabular}{c l *{5}{>{\raggedleft\arraybackslash}p{0.8cm}}} 
\toprule
 & \textbf{Model} & \textbf{Prec} & \textbf{Rec} & \textbf{F1} & \textbf{Accr} & \textbf{AUC} \\
\midrule
\multicolumn{7}{l}{\textbf{Pre-pandemic}} \\
\cmidrule(r){2-2}
 & Baseline & 0.269 & 0.500 & 0.350 & 0.539 & 0.500 \\
 & Decision Tree & 0.654 & 0.653 & 0.652 & 0.655 & 0.653 \\
 & Na\"ive Bayes & 0.697 & 0.675 & 0.673 & 0.687 & 0.749 \\
 & MLP & 0.706 & 0.706 & 0.705 & 0.707 & 0.802 \\
 & SVM & 0.737 & 0.736 & 0.736 & 0.738 & 0.812 \\
 & Logistic Regression & \textbf{0.753} & \textbf{0.749} & \textbf{0.749} & \textbf{0.753} & \textbf{0.838} \\
\midrule
\multicolumn{7}{l}{\textbf{During-pandemic}} \\
\cmidrule(r){2-2}
 & Baseline & 0.265 & 0.500 & 0.346 & 0.530 & 0.500 \\
 & Decision Tree & 0.651 & 0.651 & 0.651 & 0.654 & 0.651 \\
 & Na\"ive Bayes & 0.706 & 0.701 & 0.691 & 0.693 & 0.764 \\
 & MLP & 0.713 & 0.711 & 0.710 & 0.713 & 0.788 \\
 & SVM & 0.752 & 0.751 & 0.750 & 0.752 & 0.835 \\
 & Logistic Regression & \textbf{0.764} & \textbf{0.766} & \textbf{0.764} & \textbf{0.765} & \textbf{0.845} \\
\bottomrule
\end{tabular}%
}
\caption{Average metrics of the pre-pandemic-model and during-pandemic-model in k-fold (k=10) cross-validation.}
\label{tab:modelperformance1}
\end{table}

We analyze the misclassified posts to isolate factors contributing to misclassifications. We sample several posts from test sets with the predicted labels by pre-pandemic and during-pandemic models. We find several false positive posts containing the condemnation of discriminatory behavior. For example, \textit{``The Supreme Court ruled that companies do not have a right to discriminate against LGBTQ people in the workplace. This historic decision holds firm that LGBTQ people are, and should be, protected from discrimination under federal law.'' } This post is annotated as a non-minority stress post, suggesting the advancement of LGBTQ rights. Since there are implicit references to past discrimination behaviors in employment for the LGBTQ population, the models fail to distinguish it from minority stress posts. 
%We also observe that some posts containing violent behaviors are under the false positive category, e.g., \textit{``Police are describing the deaths of a young lesbian couple as murder/suicide, after their bodies were found in a hotel room. The pair were attending a family event the night before, when a jealous argument apparently escalated to violence.''} This post indicates the violent action towards LGBTQ people. However, it does not explicitly indicate such violent behaviors triggered by the LGBTQ identity. 
We notice some false negatives when posts mention minority stress without further details, such as: \textit{`` It's sad that we as LGBTQ Americans still cannot help save our own neighbors' lives! I feel as though it is plain and simple discrimination!''}. 
%False negative cases may also occur in a hypothetical situation. For instance, \textit{``Yes, We live in a Cisgender Heteronormative Society. If you don't fit in to the right box. If you are not a White Man, death and doom are for you. You don't want none"}

%

\begin{table}[!t]
\centering
\resizebox{0.44\textwidth}{!}{%
\begin{tabular}{c l *{5}{>{\raggedleft\arraybackslash}p{0.8cm}}} 
\toprule
 & \textbf{Features} & \textbf{Prec} & \textbf{Rec} & \textbf{F1} & \textbf{Accr} & \textbf{AUC} \\
\midrule
\multicolumn{7}{l}{\textbf{Pre-pandemic}} \\
\cmidrule(r){2-2}
 & Full & \textbf{0.753} & \textbf{0.749} & \textbf{0.749} & \textbf{0.753} & \textbf{0.838} \\
 & -Embeddings & 0.746 & 0.742 & 0.742 & 0.746 & 0.832 \\
 & -Sentiment & 0.735 & 0.732 & 0.732 & 0.735 & 0.816 \\
 & -LIWC & 0.737 & 0.735 & 0.735 & 0.738 & 0.824 \\
 & -n-grams & 0.738 & 0.736 & 0.736 & 0.739 & 0.824 \\ 
\midrule
\multicolumn{7}{l}{\textbf{During-pandemic}} \\
\cmidrule(r){2-2}
 & Full & \textbf{0.764} & \textbf{0.766} & \textbf{0.764} & \textbf{0.765} & \textbf{0.845} \\
 & -Embeddings & 0.757 & 0.759 & 0.757 & 0.759 & 0.842 \\
 & -Sentiment & 0.756 & 0.758 & 0.756 & 0.758 & 0.834 \\
 & -LIWC & 0.750 & 0.751 & 0.749 & 0.750 & 0.833 \\
 & -n-grams & 0.726 & 0.727 & 0.726 & 0.728 & 0.825 \\
\bottomrule
\end{tabular}%
}
\caption{Logistic regression model performance of the pre-pandemic-model and during-pandemic-model across different feature sets.}
\label{tab:modelperformance2}
\end{table}

\begin{table}[tb]
\centering

\begin{tabular}{l l c c} 
\toprule
 & \textbf{Feature} & \textbf{Pre} & \textbf{$\Delta$} \\ 
\midrule
\multirow{3}{*}{Minority stress} & cause & 3 & 593 $\downarrow$ \\ 
 & certain & 6 & 220 $\downarrow$ \\ 
 & family & 7 & 398 $\downarrow$ \\ 
\midrule
\multirow{3}{*}{Non-minority stress} & number & MR & 498 $\uparrow$\\ 
 & discrepancy & MR-13 & 437$\uparrow$ \\ 
 & perception & MR-15 & 521$\uparrow$ \\ 
\midrule[\heavyrulewidth]
 & \textbf{Feature} & \textbf{During} & \textbf{$\Delta$} \\ 
\midrule
\multirow{3}{*}{Minority stress} & anger & 2 & 40 $\downarrow$ \\ 
 & inhibition & 24 & 22 $\downarrow$ \\ 
 & see & 45 & 509 $\downarrow$ \\ 
\midrule
\multirow{3}{*}{Non-minority stress} & leisure & MR-1 & 96 $\uparrow$ \\ 
 & cause & MR-8 & 593 $\uparrow$ \\ 
 & posEmo & MR-11 & 497 $\uparrow$ \\
\bottomrule
\end{tabular}
\caption{Top three LIWC features for the best-performing pre-pandemic and during-pandemic model with its ranking among 607 features. MR indicates the Maximum Rank (= 607). The $\Delta$ indicates the change of rank from one model to another with its change direction. \label{tab:liwcFeature}}
\end{table}

\subsection{RQ2: Linguistic Differences of Minority Stress}
To answer RQ2, we examine the most important features of the best minority stress classifiers. We explore which features in tweets contribute to identifying the difference between minority stress in posts before and during COVID-19. These features indicate that the models consider them to be associated with the labels of ``minority stress'' assigned by human annotators and provide further insights into inferring minority stress from tweets. 

%Looking at the topmost statistically significant features, we observe that for both pre-pandemic-model and during-pandemic model.
Tables \ref{tab:liwcFeature} and \ref{tab:ngramFreature} report the top three LIWC features and top five n-gram features, indicating minority stress tweets and non-minority stress tweets with their ranking among all features. MR indicates the maximum rank (607) and $\Delta$ indicates the change of rank with its change direction. For example, feature \textit{number} with the rank MR in the pre-pandemic model indicates that its rank is 607, which is the most important feature to classify non-minority stress tweets.

In table \ref{tab:liwcFeature}, we observe that cognitive attributes, including \textit{causation}, \textit{certainty} and \textit{inhibition}, contribute significantly to the classification. \textit{Cognitive attribute} in LIWC includes words such as ``cause'', ``know'', and ``ought''. Cognition ability is linked to re-structure and understanding the emotional experience. Previous studies suggest that increasing the use of cognitive words is linked to health improvement with repeated writing and cognition ability may be essential to adaptation to traumatic or stressful experiences~\cite{klein2001expressive}. This is also consistent with our annotated dataset. For example, \textit{``It took me more than a decade to embrace my homosexuality and that it can \textbf{never} be \textbf{changed}. Even with that knowledge, I am still depressed \textit{because} sexuality is \textbf{completely} useless to me as an introvert depressed social anxiety sufferer.''} Cognitive words, including \textit{discrepancy} and \textit{cause}, also contribute to identify the non-minority stress tweets. This might be because some traumatic experiences are caused by their stigmatized identities but other heartbreaking events. Additionally, we find that the during-pandemic model heavily relies on \textit{anger}, which contains words describing violent actions or angry emotions (e.g., ``hate'', ''kill'', and ``annoyed''). For instance, \textit{``I didn't want others to assume i was gay then try to harm me or figure out where I live for future \textbf{attack}''}. 

We find many discriminating n-grams that are significant features, such as \textit{anti/anti trans}, \textit{discrimination}, \textit{discrimination homophobia}. Additionally, we also observe that the n-gram \textit{lgbtq rights} frequently occurs when complaining about Trump's statements about LGBTQ. For example, \textit{``Since Donald Trump took office, his government has attacked LGBTQ advancement. Turnout today will determine the future of \textbf{LGBTQ rights}. Please VOTE''}. One interesting observation is that the pre-pandemic classifier uses \textit{anti trans} and \textit{discrimination homophobia} to predict non-minority stress tweets. This can be explained by non-minority tweets in our dataset, which use these words as hashtags to advocate advancing LGBTQ rights.

\begin{table}[tb!]
\centering
\resizebox{0.45\textwidth}{!}{%
\begin{tabular}{l l c c} 
\toprule
 & \textbf{Feature} & \textbf{Pre} & \textbf{$\Delta$} \\ 
\midrule
\multirow{5}{*}{Minority stress} & take & 1 & 548 $\downarrow$ \\ 
 & get & 2 & 87 $\downarrow$ \\ 
 & lgbtq rights & 4 & 504 $\downarrow$ \\ 
 & anti & 5 & 177 $\downarrow$ \\ 
 & discrimination & 8 & 3 $\downarrow$ \\ 
\midrule
\multirow{5}{*}{Non-minority stress} & support & MR-1 & 456 $\uparrow$ \\ 
 & right & MR-3 & 38 $\uparrow$ \\ 
 & rt\footnote{rt refers to a re-posted tweet.} & MR-4 & 0 $-$\\ 
 & rights in~ & MR-5 & 497 $\uparrow$ \\ 
 & important & MR-6 & 192 $\uparrow$ \\ 
\midrule[\heavyrulewidth]
 & \textbf{Feature} & \textbf{During} & \textbf{$\Delta$} \\ 
\midrule
\multirow{5}{*}{\begin{tabular}[c]{@{}l@{}}\\Minority stress\end{tabular}} & being & 1 & 62 $\downarrow$ \\ 
 & anti trans & 3 & 494 $\downarrow$ \\ 
 & \begin{tabular}[c]{@{}l@{}}discrimination \\homophobia\end{tabular} & 4 & 468 $\downarrow$ \\ 
 & 2020 & 5 & 216 $\downarrow$\\ 
 & allow & 6 & 236 $\downarrow$ \\ 
\midrule
\multirow{5}{*}{Non-minority stress} & \begin{tabular}[c]{@{}l@{}}bigotry \\discrimination\end{tabular} & MR & 38 $\uparrow$ \\ 
 & music & MR-2 & 356 $\uparrow$ \\ 
 & for & MR-3 & 235 $\uparrow$ \\ 
 & rt & MR-4 & 0 $-$ \\ 
 & all the & MR-5 & 199 $\uparrow$ \\
\bottomrule
\end{tabular}%
}
\caption{Top five n-gram features for the best-performing pre-pandemic and during-pandemic model with its ranking among 607 features. MR indicates the Maximum Rank (= 607). The $\Delta$ indicates the change of rank from one model to another with its change direction. \label{tab:ngramFreature}}
\end{table}

\subsection{RQ3: Linguistic Analysis of Matched Result}
\begin{figure}
\subfloat[\label{sfig:cohend}]{%
  \includegraphics[width=.49\linewidth]{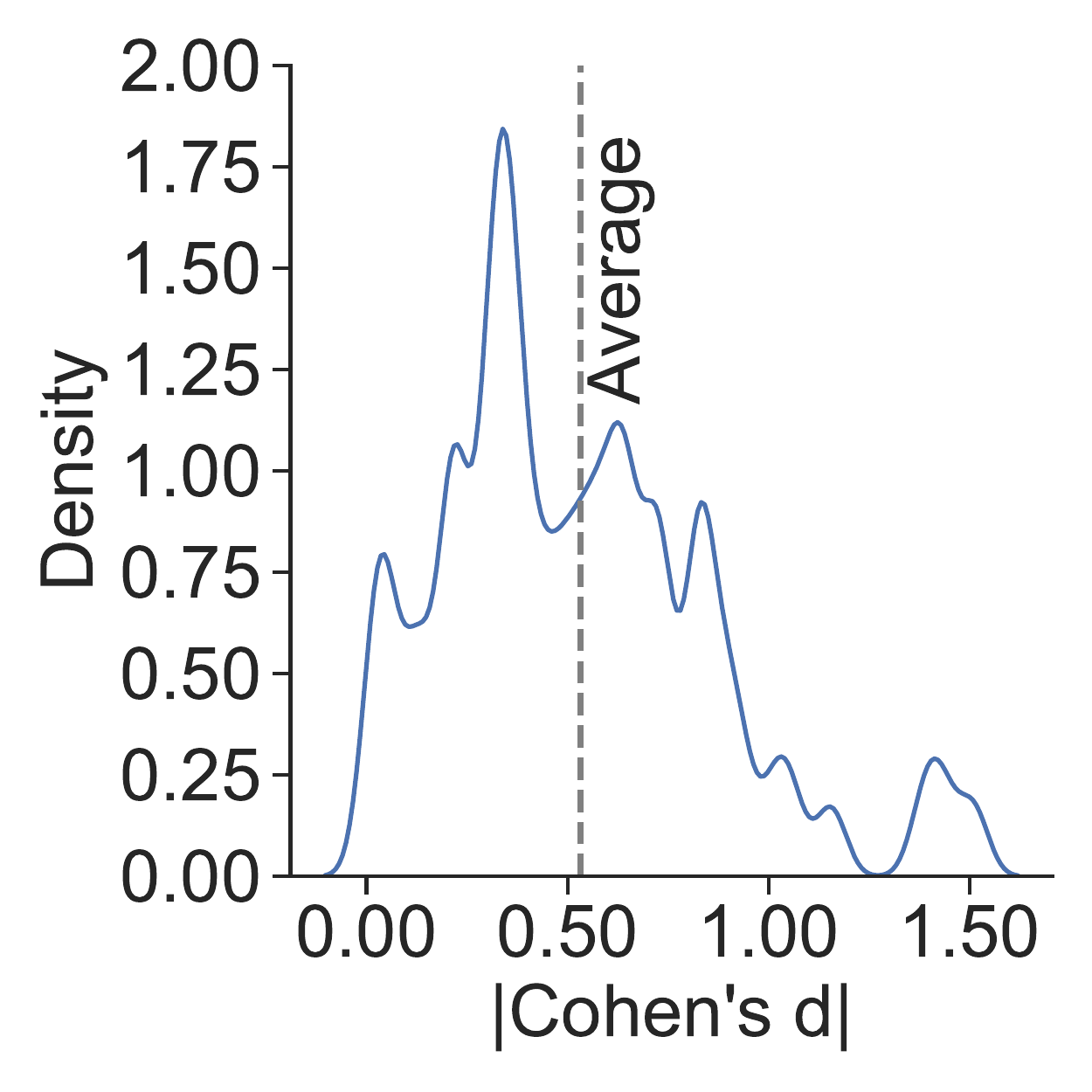}%
}
\subfloat[\label{sfig:liwcRTE}]{%
  \includegraphics[width=.50\linewidth]{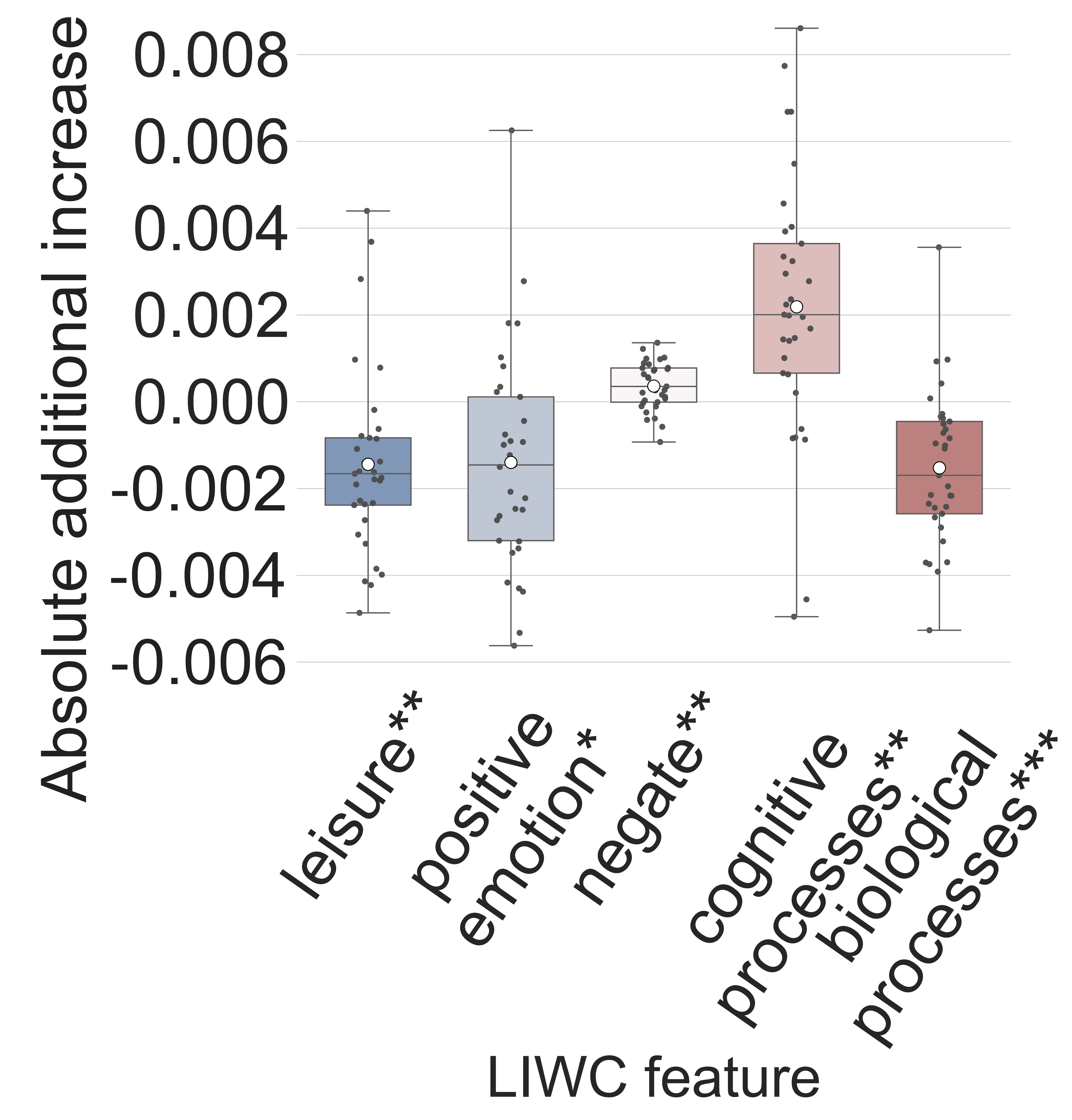}%
}
\caption{(a) Distribution of effect size magnitude in the outcome change between LGBTQ and Control users; 
(b) Distribution of treatment effect across all the LGBTQ users. Statistical significance is conducted using Welch's t-test and $p$-values are reported after Bonferroni correction (*$p < 0.05$, **$p < 0.01$, ***$p < 0.001$).}
\label{fig:liwcRTE}
\end{figure}

To investigate if the COVID-19 pandemic has a statistically significant effect on LGBTQ users, we estimate the effect size (Cohen's $d$) in LIWC word usage changes between the LGBTQ and the control group per stratum. As shown in Figure \ref{sfig:cohend}, we find that the average Cohen's $d$ in the outcome changes across all valid strata is 0.53, which illustrates medium to large differences in LIWC word usage. A Welch's t-test reveals the statistical significance of these differences ($ t\in [-5.64, 3.38], p < 0.05$), which confirms the significant changes in these outcomes.

We analyze the effect of the pandemic. Figure \ref{sfig:liwcRTE} shows the distribution of the treatment effect across all strata. The average treatment effects greater than 0 indicate a greater increase in word usage following the impact of the pandemic. We observe that the pandemic causes greater increases for the LGBTQ group on the outcomes of \textit{negation} ($ t = 3.38, p < 0.01$) and \textit{cognitive processes} ($ t = 3.38, p < 0.01$). \textit{Negation} is a category in LIWC containing ``no'', ``not'', ``never'', and so on. One explanation is that because of the disproportionate impact of the pandemic on LGBTQ groups, LGBTQ users use more cognitive words to express their traumatic experiences. We observe the average treatment effect of \textit{positive emotion} ($ t = 1.21, p < 0.05$) is less than 0, which is translated to worsened observable attribute in the usage of positive words, such as ``live'', ``nice'', and ``sweet''. We find the worse observable attributes of ($ t = 1.21, p < 0.05$), \textit{leisure} ($ t = -4.11, p < 0.01$) and \textit{biological processes} ($ t = -5.64, p < 0.001$). The \textit{leisure} in LIWC contains words related to lifestyles, such as ``cook'', ``chat'', and ``movie'', and the \textit{biological processes} contains ``eat'', ``clinic'', ``horny'', and so on.
%For the the average treatment effects that are smaller than 0, including \textit{positive emotion} ($ t = 1.21, p < 0.05$), \textit{leisure} ($ t = -4.11, p < 0.01$) and \textit{biological processes} ($ t = -5.64, p < 0.001$), the outcomes are translated to worsened observable attributes caused by the pandemic. 
Interestingly, there is no significant difference in the usage of \textit{negative emotion}, which suggests that the pandemic has no disproportionate impact for the LGBTQ group on that.

%The average treatment effect across these symptomatic LIWC features is 0.000213, and the standard derivation is 0.001046. It can be seen from the average treatment effect that the COVID-19 pandemic leads to a disproportionate impact for the LGBTQ group. All negative outcomes, including negative emotion, anxiety, sadness, show an absolute increase after the breakout of the COVID-19 pandemic. We also find that cognitive processes such as \textit{tentativeness}, \textit{causation}, \textit{certainty} have the largest absolute treatment effect. One explanation is that because of the disproportionate impact of the pandemic for the LGBTQ groups, LGBTQ users use more cognitive words to express their traumatic experience. 

\section{Conclusion and Discussion}
\subsubsection{Summary:} In this work, we develop theory-driven computational approaches to quantify minority stress from social media data. Our classifiers demonstrate strong capabilities to detect minority stress from tweets. Following this, we investigate the differences in language-related attributes of minority stress before and during the COVID-19 pandemic. We observe the increased prominence of anger words as predictive attributes of minority stress during the pandemic. To explain the differential impact on their emotions, we also contrast the linguistic difference in tweets between the LGBTQ population and the general population during the pandemic by adopting a propensity score-based matching approach. Our findings reveal that during the pandemic, the LGBTQ population is faced with a disproportionate impact on their cognitive and emotional states. This is demonstrated in their worsened observable outcome in the usage of words that convey positive emotions and a greater increase in the usage of cognitive words.

\subsubsection{Implications:} 
Past research work on LGBTQ communities has acknowledged the difficulty of accessing health-related data \cite{mcdermott2013hard}. Most existing LGBTQ health-related research is limited to diagnostic reports and psychological questionnaires. We believe that the methodologies and results from our study can provide policy-makers with important health and well-being information to support evidence-based decisions for a better understanding of the development of mental health and minority stress among the LGBTQ population. The need for such targeted approaches is more pertinent than ever in times of a global pandemic, as our findings also illustrate the disproportionate impact on the emotional and cognitive states of the LGBTQ population compared to the general population. We find that, on average, the LGBTQ population has a greater increase in the usage of cognitive words and uses positive words less often during the COVID-19 pandemic as compared to a group of the general population with comparable pre-pandemic behavior traits. Prior studies have shown that the mental health of individuals within the LGBTQ community is adversely affected due to the discrimination and bias they face. These pre-existing mental health conditions could make them even more vulnerable to the adverse outcomes of the pandemic, leading to threatening situations. 

We demonstrate the opportunities that machine learning techniques and social media data can offer to public health professionals. Our methods can be used to build tools allowing social media platforms to measure the relative increase of minority stress among LGBTQ individuals during public crises. This aligns with the American Counseling Association’s Code of Ethics \cite{american2014american}, which suggests that counselors need to gain knowledge, sensitivity and skill to work with diverse backgrounds. Our findings can provide counselors with sensitivity and awareness about the discrimination and stigmatization that LGBTQ individuals face to build client-centered practices. 

We believe that our research can complement the tools to develop the minority stress theory further. Broadly, several studies have discussed increased fear, anger, and trauma during the COVID-19 pandemic. Aligned with these broader findings, our work demonstrates an increased association of anger with minority stress in LGBTQ groups. Our results can be used to build tools that reveal the feelings and emotions behind the language of LGBTQ people. Understanding the social media communication from the LGBTQ population can offer a new lens onto minority stress at the population level, which can be utilized to verify the existing research on minority stress or identify the under-observed variables of minority stressors.

\subsubsection{Limitations:} Our work is grounded in Meyer's theory of minority stress. Although it is widely used for various empirical investigations on the health of the minority population, this framework has its own limitations. As acknowledged by \citeauthor{meyer2003prejudice} (\citeyear{meyer2003prejudice}), there is a necessity to expand this framework and distinguish between different stigmatized statuses. Future work could dig deeper into this issue by adding different psychological scales for different groups. 

Another limitation of our study is selection biases.  We search Twitter users using hashtags, and collect self-disclosed identity information in their bios. This is likely to be influenced by self-selection bias as we can only collect data from those who expressed themselves on social media. This is especially true when considering that LGBTQ people have long been exposed to stigmatization and prejudice. Similarly, this research exclusively concentrates on Twitter, and this could lead to biased or insufficient viewpoints. Consequently, our findings may not be generalized to other online communities.

A further note is that we only use features inside the tweets to measure minority stress. Other features, such as frequency of posting and author metadata, could be incorporated to analyze the minority stress comprehensively. Another research direction is to explore posts on minority stress across different social media platforms to build accurate, comprehensive, and up-to-date feature sets and gain a better and more thorough understanding of this topic.

Lastly, our analysis does not differentiate results across subcategories of minority stress due to the difficulty of extracting the context from a single tweet. Future works can analyze minority stress from the conversation level on Twitter to better understand minority stress. 

%\subsection{Privacy Policy and Ethics}
%With the increased use of social media as a source to analyze health information,  users privacy and confidentiality has %become a growing concerns. 

\section{Acknowledgements}
The calculations presented above were performed using computer resources within the Aalto University School of Science “Science-IT” project. This work has been supported by Helsinki Institute for Information Technology HIIT.

\section{Supplementary Material}

\textbf{LGBTQ Related Hashtag List:} \textit{ lgbtq, gay, pride, lgbtqia, pridemonth, love, queer, trans, pridemonth2021, pride2021, mmromance, loveislove, gaymer, lesbian, gayromance, nonbinary, transgender, hiv, gaydar, equality, bisexual, gayrights, paranormal, comingout, pansexual, transrightsarehumanrights, transisbeautiful, homophobia, lgbtqpride, rainbow, lgbtpride, kpopstans, lgbt, clubsilver, gaytravel, lgbttravel, instagay, gayboy, gayart, dragqueen, gaytwitter, gayguy, bara, yaoi, bear, sexy, gayman, queen, gayuk, furry, gaylove, bi, queerart, grindr, queenofpop, gaybear, gayfiction, gaymen, scruff, queerfilmrec.}

\textbf{Non-LGBTQ Related Hashtag List:} \textit{sunday, birthday, yelp, covid19, florida, model, smile, losangeles, fitness, coronavirus, dogsofinstagram, tiktok, family, makeup, faith, catsofinstagram, blacklivesmatter, newprofilepic, style, foodporn, nowplaying, chicago, summer, grateful, truth, travel, fun, life, photography, nyc, foodie, picoftheday, latergram, nofilter, fashion, artist, halloween, friends, wynonnaearp, music, art, shopmycloset, beach, neasummit, disney, vote, podcast, happy, texas, miami, food, sundayfunday, beautiful, nature, blessed, home, inspiration, usa, flowers, happybirthday, instagram, california, selfie, soundcloud, vacation, vegan, poshmark, twitch, sunset, beauty, photooftheday, christmas, goodmorning, repost, realestate, motivation, instagood, happiness, newyork.}

\bibliography{aaai23}

\end{document}